\documentclass[aps,pra,twocolumn,groupedaddress,showpacs,showkeys]{revtex4-1}
\usepackage{amsmath}
\usepackage{mathtools}
\begin{document}

%\preprint{}

\title{Nonlinear Stokes Mueller Polarimetry}


\author{Masood Samim, Serguei Krouglov, Virginijus Barzda}
\email[]{virgis.barzda@utoronto.ca}
%\thanks{}
\altaffiliation{Department of Chemical and Physical Sciences,University of Toronto Mississauga, 3359 Mississauga Road North, Mississauga, Ontario L5L1C6, Canada}
\affiliation{Department of Physics and Institute for Optical Sciences, University of Toronto, 60 St. George Street, Toronto, Ontario M5S1A7, Canada}


\date{\today}

\begin{abstract}
The Stokes Mueller polarimetry is generalized to include nonlinear optical processes such as second- and third-harmonic generation, sum- and difference-frequency generations. The overall algebraic form of the polarimetry is preserved, where the incoming and outgoing radiations are represented by column vectors and the intervening medium is represented by a matrix. Expressions for the generalized nonlinear Stokes vector and the Mueller matrix are provided in terms of coherency and correlation matrices, expanded by higher-dimensional analogues of Pauli matrices. In all cases, the outgoing radiation is represented by the conventional $4\times 1$ Stokes vector, while dimensions of the incoming radiation Stokes vector and  Mueller matrix depend on the order of the process being examined. In addition, relation between nonlinear susceptibilities and the measured Mueller matrices are explicitly provided. Finally, the approach of combining linear and nonlinear optical elements is discussed within the context of polarimetry.
\end{abstract}


\maketitle

\section{Introduction}
Polarimetry techniques employ the measurement of polarization state of optical response from the sample for a defined polarization of incoming radiation. Linear optical polarimetry is a well established measurement technique that found applications in different research fields including material science and biomedical imaging~\cite{Bickel1976_Appli, Bickel1981_Polar, Yao1999_Two-d, Jiao2000_Depth, Ghosh2008_Muell, Ghosh2009_Muell, Wood2009_Proof}. Polarimetry can also be employed for nonlinear optical techniques, such as second-harmonic generation (SHG), third-harmonic generation (THG), coherent anti-Stokes Raman scattering (CARS)~\cite{Campagnola1999_High-,Campagnola2006_Secon,Nucciotti2010_Probi,Bancelin2014_Deter,Boulesteix2004_Secon,Samim2014_Secon,Freund1986_Secon,Barzda2004_Secon}.

In an optical setup the polarization-dependent interaction of light with matter can be described using Stokes Mueller, Poincar\'e or Jones formalism~\cite{Shurcliff1962_Polar,Kliger1990_Polar,Azzam1977_Ellip}. Each formalism has unique advantages conveniently applicable for different circumstance. In the linear Stokes Mueller formalism, the light is represented by a four-element Stokes vector, and its interaction with matter is represented by a $4\times4$ Mueller matrix. The Stokes vector can describe partially- or completely-polarized light, and operates with intensities, which are real numbers, and thus, observables in an experiment. On the other hand, Jones formalism is used to describe purely polarized states retaining the phase relations of the electric fields and requires working with complex variables.

Recently, attempts have been made to deal with the nonlinear polarization measurements in a linear fashion. For example SHG signal from samples have been characterized by a $4\times1$ Stokes vector as well as for the incoming fundamental beam radiation~\cite{Lien2013_Preci,Mazumder2014_Revea}. However, the characterization of the sample remain unresolved, mainly because of the nonlinear relationship between the incoming and outgoing radiations. In ellipsometry, for two-photon processes some nonlinear relationships have been derived by using a quantum-mechanical framework and a Jones Stokes approach~\cite{Shi1994_Gener,Begue2009_Nonli2}. These recent efforts demonstrate the need for a unifying and general framework for nonlinear optical Stoke Mueller polarimetry.

Here, we develop the theoretical framework for nonlinear optical polarimetry by using the classical description of electric fields, nonlinear susceptibilities and optical radiations polarizations. The polarization state of light as well as the response of a material is described with real-valued parameters. In our approach to the multi-photon polarimetry, the Jones and Stokes Mueller formalism is analogous to the conventional linear polarimetry. The Jones formalism can be used to describe nonlinear light-matter interaction using higher-order susceptibilities and pure polarization states. However, often media, including biological tissue, are a highly heterogeneous scattering materials; therefore, there is an additional benefit to employ Stokes Mueller formalism for analyses of nonlinear optical responses. Additionally, the linear polarimetry technique has extensive and comprehensive formulations for describing a measurement system that may be applicable to a nonlinear polarimetry experiment. For example, the degree of polarization of an optical radiation is a useful parameter to quantify the extent of coherent light contribution to the radiation. In addition, various filtering mechanisms exist to separate the polarized components of a radiation from non-coherent contributions~\cite{Cloude1986_Group,Gil2007_Polar}. Analogous approaches exist in the nonlinear Stokes Mueller polarimetry as we will show in this paper.
\section{Theory of Nonlinear Stokes Mueller Polarimetry}
The general nonlinear Stokes Mueller equation, describing the relationship between the generated nonlinear signal radiation, the nonlinear properties of the media, and the incoming radiations can be written as follows:
%%%%% Eq: Stokes Poloarimetry  %%%%%
\begin{equation}\label{eq:NLStokeMueller}
s' (\omega_\sigma)= \mathcal{M}^{(n)}S(\omega_1,\omega_2,\cdots,\omega_n)
\end{equation}
where $s'$ is the Stokes vector describing the generated radiation at $\omega_\sigma$ frequency and prime signifies the measured outgoing signal, $\mathcal{M}^{(n)}$ is the nonlinear Mueller matrix describing an $n^\text{th}$ order light-matter interaction, while $S$ is a vector representing the incoming electric fields that generate the light via nonlinear interactions. Henceforth, the $s'$ and $S$ are called the polarization state vectors for outgoing and incoming radiations, respectively.

The left hand side is the outgoing radiation, which once generated is simply represented with an ordinary $4 \times 1$ Stokes vector. The right hand side variables at a more basic level each represent two physical quantities: the nonlinear susceptibility, which is directly related to the structure of the material, and nonlinearly interacting electric fields from the incoming light radiation. At this level the two key variables are: $\psi$, or the state function of fields that interact to produce a particular nonlinear phenomenon; and $\chi^{(n)}$, the nonlinear susceptibility matrix that represents the material in the context of polarimetry. The polarization density for nonlinear optical interactions can be stated as follows:
\begin{equation}\label{eq:NLPolChiEField}
P_i^{(n)} = \chi_{ijk\cdots m}^{(n)}{ E_j}{ E_k}\cdots { E_m}= {\chi_{iA}^{(n)}}{\psi_A^{(n)}}
\end{equation}
where (Einstein) summation is assumed over the repeated indices. The first index for $\chi$ represents the orientation of the outgoing polarization and the remaining indices represent the direction of polarization for incoming electric fields. The relation between the index $A$ and $j,k,\dots,m$ is specific for a given nonlinear process. Essentially, for an $n^\text{th}$ order nonlinear optical phenomena $A$ runs from $1$ to $n+1$, and $i$ represents the two orthogonal vectors expanding the plane of polarization perpendicular to the direction of light propagation~\footnote{Thus, any matrix notation of $\chi_{iA}^{(n)}$ (i.e. with two indices) represent the contracted notation of $n^\text{th}$ order nonlinear susceptibilities.}.

The Stokes vector can be measured using the light intensity, and the nonlinear outgoing intensity depends on the susceptibility and the interacting electric fields according to the following equation:
\begin{equation}\label{eq:NLIntensity}
I \propto  P_iP_i^*  \propto {\chi _{iA}^{}}\chi _{iB}^* {\psi _A}\psi _B^ *
\end{equation}
Thus, we see that Stokes and Mueller notations are composed of products of electric fields vectors, and products of susceptibilities components, respectively.
At the level of individual electric fields, the outgoing field, denoted by the state vector $\Phi'$, is related to the products of incoming nonlinear electric fields, denoted by the state vector $\psi^{(n)}$, which interacts with the nonlinear susceptibility that is denoted by $\chi^{(n)}$:
\begin{equation}\label{eq:PhiChiPsi}
\boxed{\Phi'(\omega_\sigma)=\chi^{(n)} \psi^{(n)}(\omega_1,\omega_2,\cdots,\omega_n)}
\end{equation}
In this framework, each component of vector $\Phi'$ of the generated electric field is proportional to the polarization density, and it depends on the susceptibility tensor components as well as on the state of the laser polarization that has $n+1$ components (see Eq.~\ref{eq:NLPolChiEField}). The state vector for the nonlinear combination of electric fields in the incoming radiation is:
\begin{equation}
\psi^{(n)}(\omega_1,\omega_2,\cdots,\omega_n)=
\left(
\begin{array}{c}
\psi^{(n)}_1\\
\psi^{(n)}_2\\
\vdots\\
\psi^{(n)}_{n+1}\\
\end{array}
\right)
\end{equation}
Each element of the state vector $\psi^{(n)}_A$ ($A = 1, \cdots, n+1$) is an $n^\text{th}$ order function of one or more electric fields oscillations at particular frequencies. 
\subsection{Outgoing Radiation Stokes Vector}
The Stokes vector $s'$  for the outgoing electric field $E(\omega_\sigma)$ is characterized by a $4\times 1$ vector just as in the case for conventional Stokes vector. Let ${C}'(\omega_\sigma)=\left<\Phi'(\omega_\sigma) \cdot \Phi'^{\dag}(\omega_\sigma)\right>$ be the coherency matrix composed from the dyad of $\Phi$, where  $\Phi(\omega_\sigma)$ is the state (or simply the electric field) vector of the outgoing beam. $\left< \cdot \right>$ signifies a time average over an interval long enough to make the time-averaging independent of the interval and fluctuations. Then, in terms of its coherency matrix and Pauli matrices the outgoing field Stokes vector is~\cite{Samim2015_Doubl,Azzam1977_Ellip}:
%%%%% Eq: Stokes Poloarimetry  %%%%%
\begin{equation}\label{eq:StokeNLOutgoing}
{s'_t } = {\rm{Tr}}\left( {{C}'{\tau_t^{}}} \right)= {{C}'_{ab}}{\left( {\tau_t^{}} \right)_{ba}} = \left<{\Phi'_a}\Phi_b'^*\right>{\left( {\tau_t^{}} \right)_{ba}} = \left<{\Phi '^\dag }{\tau_t }\Phi '\right>
\end{equation}
where $a$ and $b$ each run from 1 to 2, representing the orthogonal outgoing polarization orientations perpendicular to the propagation direction. $\tau_t$ ($t=0...3$) denotes the $2\times 2$ identity and Pauli matrices:
%%%%% Eq: Pauli Matrices  %%%%%
\begin{equation}\label{eq:PauliXZ}
\begin{array}{*{20}{c}}
{{\tau}_0 = \left( {\begin{array}{*{20}{c}}
		1&0\\
		0&1
		\end{array}} \right)} & {{\tau}_1 = \left( {\begin{array}{*{20}{c}}
		1&0\\
		0&-1
		\end{array}} \right)}\\
{{\tau}_2 = \left( {\begin{array}{*{20}{c}}
		0&1\\
		1&0
		\end{array}} \right)}&{{\tau}_3 = \left( {\begin{array}{*{20}{c}}
		0&-i\\
		i&0
		\end{array}} \right)}
\end{array}
\end{equation}

The so-called degree of polarization ($dop$) is defined as~\cite{Born1999_Princ}:
\begin{equation}
	dop = \sqrt{{s'_1}^2+{s'_1}^2+{s'_1}^2}/s'_0
\end{equation}
\subsection{Real-valued Vector for Incoming Radiation}
The nonlinear electric fields have a $(n+1)\times(n+1)$ coherency matrix which is defined as:
\begin{equation}\label{eq:CoherencyGen}
\begin{split}
\rho^{(n)}&(\omega_1,\omega_2,\cdots,\omega_n)=\left<\psi^{(n)}\cdot \psi^{(n)\dag}\right>\\
&= 
\begin{pmatrix}
\left<\psi^{(n)}_1 \psi^{(n)*}_1\right>     & \cdots & \left<\psi^{(n)}_1 \psi^{(n)*}_{n+1}\right>\\
\vdots					       & \ddots & \vdots\\
\left<\psi^{(n)}_{n+1} \psi^{(n)*}_1\right> & \cdots & \left<\psi^{(n)}_{n+1} \psi^{(n)*}_{n+1}\right>\\
\end{pmatrix}
\end{split}
\end{equation}
For nonlinear interaction of electric fields the coefficients of expansion for the coherency matrix forms a real-valued vector similar to the Stokes vector. The nonlinear coherency matrix can be expanded by basis that have higher dimensions than the Pauli's matrices. Leaving aside the details of the dimension for now, and simply denoting this set as $\eta$, the nonlinear vector can be written as:
\begin{equation}\label{eq:StokeNLIncoming}
\boxed{{S_N} = {\rm{Tr}}\left( {\rho}\, {\eta_{_N}^{}} \right) 
= \left<{\psi _A}\psi _B^* \right> {\left( {\eta_{_N}^{}} \right)_{BA}} = \left<{\psi ^\dag }{\eta_{_N}^{}}\psi \right>}
\end{equation}
where $N =1,\cdots, (n+1)^2$ for each element of the nonlinear vector representing the $n^\text{th}$ order electric fields. The $\eta$ matrices similar to Pauli's expand higher dimension states. A subset of properties of $\eta$ matrices essential for deriving an $n^\text{th}$ order process is:
\begin{enumerate}
	\item They are square matrices with dimension $(n+1)\times(n+1)$.
	\item They are hermitian: $\eta^\dag=\eta$.
	\item There are $(n+1)^2$ of $\eta$ matrices which form the basis.
	\item Obey the orthogonality $\text{Tr}(\eta_\mu\eta_\nu)=c_\eta\delta_{\mu\nu}$ where $c_\eta$ is a constant and real number, and $\delta_{\mu\nu}$ is the Kronecker delta ($\delta_{\mu\nu}=1$ when $\mu=\nu$, and 0 otherwise).
\end{enumerate}
The constant $c_\eta$ can be chosen to be the same and equal to 2 for any order of interaction, similar to the linear case (for Pauli matrices $\text{Tr}(\tau_\mu\tau_\nu)=2\delta_{\mu\nu}$). The recipe for finding these matrices is given in the section $\eta$ \textit{Matrices for Nonlinear Polarimetry}.

Similar to the linear Stokes parameters the nonlinear vector obeys the following relation~\footnote{For any vector $\psi$ of length $n+1$, let $\rho=\psi\otimes\psi^\dag$: then $n\,{\left[\text{Tr}(\eta_{1}\,\rho)\right]}^2=\sum\limits_{N=2}^{(n+1)^2}{{\left[\text{Tr}(\eta_{N}^{}\,\rho)\right]}^2}$. The set of $\eta$ matrices are defined in Section~\ref{sec:eta}}:	
\begin{equation}\label{eq:S1Gen}
n\,S_1^2 \ge \sum\limits_{N = 2}^{{(n+1)}^2} {S_N^2} 
\end{equation}	
where the equality is valid for the purely polarized state. Therefore, it is helpful to use the degree of polarization (DOP) parameter to characterize the fundamental radiation using the nonlinear vector:
\begin{equation}\label{eq:DOPGen}
DOP(\omega_1,\omega_2,\cdots,\omega_n) = \sqrt {\sum\limits_{N = 2}^{{(n+1)}^2} {S_N^2} /nS_1^2}
\end{equation}
where $DOP$ ranges from 0 to 1 for unpolarized to fully polarized fundamental radiation, respectively.
\subsection{Real-valued Matrix for Intervening Medium}\label{sec:NLMueller}
By substituting linear and nonlinear Stokes vector expressions (Eq.~\ref{eq:StokeNLOutgoing} and~\ref{eq:StokeNLIncoming}, respectively) into the general nonlinear polarimetry Eq.~\ref{eq:NLStokeMueller} the following expression is obtained:
\begin{equation}\label{eq:OutnInStates}
\left< {\Phi '^\dag }\,{\tau^{}_t }\,\Phi ' \right>= \mathcal{M}_{t N}^{(n)}\,\left< {\psi ^\dag }\,{\eta_{_N}^{}}\,\psi\right>
\end{equation}
In this frame, each component of the vector $\Phi$ of the generated electric field is proportional to the polarization, which depends on the susceptibility tensor components and the polarization state of the radiation of incoming nonlinear electric fields. By substituting explicit expressions of $\Phi'$ and $\Phi'^\dag$ into Eq.~\ref{eq:OutnInStates} in the elemental form:
%%%%% Eq: Stokes Poloarimetry  %%%%%
\begin{equation}\label{eq:NLStokeMuellerElments}
\left<  \chi^{(n)*}_{aA}\psi _A^*{\left( {{\tau^{}_t }} \right)_{ab}}{\chi^{(n)}_{bB}}{\psi _B} \right> = {\mathcal{M}^{(n)}_{t N}} \left< {\psi_A^*}\,({\eta_{_N}})_{AB}\,\psi^{}_B\right>
\end{equation}
where $A$ and $B = 1, \cdots, n+1$. Since Eq.~\ref{eq:NLStokeMuellerElments} is written in terms of individual elements, the state functions of the fundamental radiation can be dropped and the nonlinear Mueller matrix elements $\mathcal{M}_{tN}$ can be written in terms of the $n^\text{th}$ order susceptibilities as:
%%%%% Eq: Stokes Poloarimetry  %%%%%
\begin{equation}\label{eq:MuellerNL}
\chi _{aA}^*{\left( {{\tau^{}_t }} \right)_{ab}}{\chi^{}_{bB}} = {\mathcal{M}_{t N}}{\left( {{\eta_{_N}}} \right)_{AB}}
\end{equation}
Note, in Eq.~\ref{eq:MuellerNL} the signal is assumed to be from a single generator, and an ensemble of scatterers have a similar derivation, which will be shown in Section F. Multiplying both sides by ${(\eta_{_{N'}})}_{BC}$ and after summation over index $B$, and letting $A=C$:
%%%%% Eq: Stokes Poloarimetry  %%%%%
\begin{equation}\label{eq:NLStokeMuellerElmentsNoTrace}
{\textstyle{1 \over c_\eta}}\chi _{aA}^*{\left( {{\tau^{}_t }} \right)_{ab}}{\chi^{}_{bB}}{\left( {{\eta_{_N}}} \right)_{BA}} = {\mathcal{M}_{t N}} 
\end{equation}
where (Einstein) summation is implied over repeated indices (i.e. $a$, $b$, $A$ and $B$). $c_\eta$ is a real-valued constant (and can be set to equal to two as will be shown in Section E). Finally, the expression of a real-valued matrix element in terms of the susceptibilities is:
%%%%% Eq: Double Mueller  %%%%%
\begin{equation}\label{eq:MuellerNLBoxed}
\boxed{{\mathcal{M}_{t N}} = \textstyle{1 \over {c_\eta}}{\rm{Tr}}\left( {{\tau^{}_t }\, \chi \, {\eta_{_N}} \, {\chi ^\dag }} \right)}
\end{equation}
This expression has a general form and is equivalent to the linear Mueller matrix element expression if the matrices $\eta$ are replaced with Pauli matrices (from Eq.~\ref{eq:PauliXZ}). In contrast to linear Mueller matrix elements, the nonlinear $\mathcal{M}$ is composed of nonlinear susceptibilities and $\eta$ matrices of higher dimension. Note that for linear polarimetry, the transformation matrix $J$ can also be represented by the linear susceptibility $\chi^{(1)}$, in which case the only difference between linear and nonlinear Mueller matrix elements would be to replace one Pauli matrix with an $\eta$ matrix~\cite{Azzam1977_Ellip,Kim1987_Relat}. This familiar form of Mueller matrix elements can be investigated similar to the linear case. All elements of nonlinear matrix are real, a fact that leads to a very useful and a much desired expression for determining the nonlinear susceptibilities.
\subsection{Expression of Susceptibilities in Nonlinear Polarimetry}
Stokes polarimetry measures the Mueller matrix components, while nonlinear properties of the material is often described by $\chi^{(n)}$ tensor component values. Thus, the next step is to derive expressions for $\chi^{(n)}$ products in terms of $\mathcal{M}_{t N}^{(n)}$ component values. To this end, we can arrive at an equivalent conclusion by using the trace property $\text{Tr}(AB)=\text{vec}(A^T)^T \text{vec}(B)$, where $\text{vec}(A) = [a_{1,1},...,a_{s,1},a_{1,2},..,a_{s,2},...,a_{1,t},...,a_{s,t}]^T$ is the vectorization of a $s\times t$ matrix $A$ (in other words columns of a matrix are stacked below one another), and its corollary $\text{Tr}(A^TBCD^T)=\text{vec}(A)^T (D\otimes B) \text{vec} (C)$ on the (real) double Mueller elements in Eq.~\ref{eq:MuellerNLBoxed} is:
\begin{equation}\label{eq:}
\begin{split}
\mathcal{M}_{t N} = \left({\mathcal{M}_{t N}}\right)^* 
& = \left(  \textstyle{1 \over {c_\eta}}{\rm{Tr}}\left( {{\tau^{}_t }\chi {\eta_{_N}}{\chi ^\dag }} \right)\right)^*  \\
& = \textstyle{1 \over {c_\eta}}{\rm{Tr}}\left( {{\tau_t^T }\chi^* {\eta_{_N}^*}{\chi ^T }} \right)\\
&= \textstyle{1 \over {c_\eta}}\text{vec}(\tau^{}_t)^T (\chi\otimes \chi^*) \text{vec} (\eta_{_N}^*)
\end{split}
\end{equation}
where in going from the first line to the second we took advantage of the hermitian properties of the $\tau$ and $\eta$. By letting:
\begin{equation}\label{eq:VecPauli}
\mathcal{T}\equiv
\left(
\begin{matrix}
\text{vec}(\tau_0)^T\\
\text{vec}(\tau_1)^T\\
\text{vec}(\tau_2)^T\\
\text{vec}(\tau_3)^T\\
\end{matrix}
\right) =
\left(
\begin{matrix}
1&0&0&1\\
1&0&0&-1\\
0&1&1&0\\
0&\mathrm{i}&-\mathrm{i}&0\\
\end{matrix}
\right),
\end{equation}
The matrix $\mathcal{T}$ is invertible and obeys $\mathcal{T}^{-1}=\textstyle{1 \over 2} \mathcal{T}^\dag$.
By letting $H^{\dag} = [\text{vec} (\eta_1^*),\cdots,\text{vec} (\eta_N^*)]$, we arrive at: 
\begin{equation}\label{eq:MviaX}
\mathcal{M} = \mathcal{T}{\rm{X}}H^{-1}
\end{equation}
where ${\rm{X}}=\chi \otimes \chi^*$. Therefore, H should be invertible and obey $H^{-1}=\frac{1}{c_\eta}H^{\dag}$. Consequently, the susceptibility products can be easily found as:
\begin{equation}\label{eq:ChifromM}
\boxed{{\rm{X}} = \mathcal{T}^{-1}\mathcal{M}H}
\end{equation}
The relationship between the nonlinear susceptibilities in terms of Mueller matrix derived in Eq.~\ref{eq:ChifromM} is useful when the Mueller matrix is obtained by the polarimetry measurement of a sample and the explicit values for the corresponding susceptibilities are desired.

In the elemental form $\text{X}_{ij}=\frac{1}{2}\mathcal{T}^\dag _{it}\mathcal{M}_{tN}H_{Nj}$, where $i = (a-1)2+b$ and $j =(A-1)(n+1)+B$ ($a$ and $b$ = 1,2; $A$ and $B = 1,\dots,n+1$). Since, $\chi_{aA}^{}\chi_{bB}^*=|\chi_{aA}||\chi_{bB}|e^{i(\delta_{aA}-\delta_{bB})}$, then the relative phase between any two susceptibility elements $\chi_{aA}^{}$ and $\chi_{bB}^{}$ can be found according to:
\begin{equation}\label{eq:RelPhaseGen}
\begin{split}
&\delta_{aA}-\delta_{bB} 
=\Delta_{aA,bB}\\
& = \tan^{-1}\left(-\mathrm{i}\frac{\chi_{aA}^{}\chi_{bB}^*-\chi_{bB}^{}\chi_{aA}^*}{\chi_{aA}^{}\chi_{bB}^*+\chi_{bB}^{}\chi_{aA}^*}\right)
=\tan^{-1}\left(\mathrm{i}\frac{\text{X}_{kl}-\text{X}_{ij}}{\text{X}_{kl}+\text{X}_{ij}}\right)\\
 &= \tan^{-1}\left(\mathrm{i}\frac{\mathcal{T}^\dag _{kt}\mathcal{M}_{t N}H_{Nl}-\mathcal{T}^\dag _{it}\mathcal{M}_{t N}H_{Nj}}{\mathcal{T}^\dag _{kt}\mathcal{M}_{t N}H_{Nl}+\mathcal{T}^\dag _{it}\mathcal{M}_{t N}H_{Nj}}\right)  
\end{split}
\end{equation}
where $k = (b-1)2+a$ and $l =(B-1)(n+1)+A$, and summations over repeated indices are assumed. Equation~\ref{eq:RelPhaseGen} is important because it shows that by measuring the  material nonlinear matrix, and using matrices $\mathcal{T}$ in Eq.~\ref{eq:VecPauli} and $H$, the relative phase of the susceptibility elements can be obtained.

In nonlinear polarimetry studies, it is customary to characterize nonlinear optical properties of the material using susceptibility values. Therefore, Eqs.~\ref{eq:ChifromM} and~\ref{eq:VecPauli} provide a mechanism to check and compare nonlinear polarimetry investigations with similar previous studies using conventional nonlinear optics. For example, the ratio of susceptibilities for cylindrically symmetric material can be calculated for a number of biological structures.
\subsection{$\eta$ Matrices for Nonlinear Polarimetry}\label{sec:eta}
The polarization state of incoming radiation $S$ (Eq.\ref{eq:StokeNLIncoming}) as well as the matrix representing the nonlinear medium $\mathcal{M}$ (Eq.\ref{eq:MuellerNLBoxed}) require the $(n+1)\times(n+1)$ $\eta$ matrices in order to be defined from the nonlinear coherency and susceptibility matrices, respectively. The recipe for generating $\eta$ matrices has two steps: In \fbox{Step 1} the matrix $\eta''_{jk}$ is defined such that only the value of element $jk$ of the matrix  $\eta''_{jk}$ is $1$, and $0$ for all other elements (both $j$ and $k$ run from 1 to $n+1$). This creates a two dimensional set of matrices, where each element of the set is a $(n+1)\times(n+1)$ matrix. Note that the $\eta''$ are also independent basis and can expand the coherency matrix. However, they are not hermitian and therefore the resulting Stokes vector and Mueller matrix will be complex. To obtain the desired hermitian matrices for an $n^\text{th}$ order process the following relation can be used:
\begin{equation}\label{eq:GGM}
\eta'_{jk}=
\begin{dcases}
\eta''_{jk}+\eta''_{kj} \; ,   \hskip 70pt \text{if } j<k\\
\mathrm{i}(\eta''_{jk}-\eta''_{kj})\; ,   \hskip 60pt \text{if } j>k\\
\sqrt{\frac{2}{j^2+j}} \left[ \left(\sum\limits_{m=1}^{j}\eta''_{mm}\right) -j\eta''_{j+1,j+1} \right]\; ,    \\ \hskip 70pt \text{if } 1\leq k=j<(n+1)\\
\sqrt{2\over {n+1}}\mathcal{I}_{n+1}\; ,    \hskip 8pt \text{if } j=k=(n+1)
\end{dcases}
\end{equation}
where $\mathcal{I}_{n+1}$ is the $(n+1)\times (n+1)$ identity matrix. The first case (when $j<k$) the new matrices $\eta'_{jk}=\eta''_{jk}+\eta''_{kj}$ are real valued; the second case (when $j>k$), the new matrices $\eta'_{jk}=\mathrm{i}(\eta''_{jk}-\eta''_{kj})$ are complex valued and have similar nonzero elements as to their real-value counterparts in the first case. In the third case, (when $1\leq j=k< n+1$), the new matrices are diagonal and real valued. Finally, in the last case an identity matrix is used. In \fbox{Step 2} the two-dimensional $\eta'$ set is converted to a one-dimensional set of matrices\footnote{This is to simplify the indices and to conform to a Stokes Mueller notation of vector = matrix $\times$ vector. The matrices $\eta'$ can also be used directly for polarimetry, in which case there will be an additional index for the entity representing the incoming radiation as well as for the entity representing the medium.}: $\eta'_{jk}\rightarrow\eta^{}_N$. 

These matrices satisfy all the requirements as desired for expanding the nonlinear coherency matrix for the nonlinear polarimetry. In addition, the new matrices defined in Eq.~\ref{eq:GGM} ensure that $\eta$ obey: $\text{Tr}(\eta_\mu\eta_\nu)=2\delta_{\mu\nu}$. For linear polarimetry $n=2$ and $\eta$ corresponds to Pauli matrices. For second-order process $n=3$ and therefore the generated matrices are those of Gell-Mann's. For the case of three photon-polarimetry $n=4$, and there are sixteen $4\times 4$ matrices, which will be shown in a seperate manuscript~\footnote{MS, SK and VB have a manuscript under preparation titled ``\textit{Third-harmonic generation Stokes Mueller Polarimetry}"}; and so forth. A useful relations between these matrices and Stokes Mueller formalism is the following: The real-valued $\eta$ generate the Stokes vector components that depend on linear polarization, while the complex valued ones are responsible for circular components. Also, the real-valued ones are in part responsible for nonzero Mueller matrix elements, while the Mueller matrix component constructed from a complex-valued $\eta$ matrix may be zero if the involved nonlinear susceptibilities are real.

\subsection{Ensemble Representation}\label{sec:Ensemble}
In a highly scattering media such as in biological tissue, the system may not be completely coherent, and the source of the signal may be an ensemble of scatterers. Therefore, an ensemble average of individual elements with probability $p_e$ may be more appropriate to consider~\cite{Kim1987_Relat}. The outgoing nonlinear radiation resulting from an ensemble is:
\begin{equation}\label{eq:NLStokeMuellerElmentsEnsemble}
\sum_{e}^{} p_e \left<  \chi^{(n)*,e}_{aA}\psi _A^*{\left( {{\tau^{}_t }} \right)_{ab}}{\chi^{(n),e}_{bB}}{\psi^{}_B} \right> = {\mathcal{M}^{(n)}_{t N}} \left< {\psi_A^*}\,({\eta_{_N}})^{}_{AB}\,\psi^{}_B\right>
\end{equation}
Since the above equation is in the elemental form, it can be rewritten as: 
\begin{equation}
{\left( {{\tau^{}_t }} \right)_{ab}} \sum_{e}^{} (p_e \chi^{(n)*,e}_{aA}{\chi^{(n),e}_{bB}} )\left<\psi _A^*{\psi^{}_B} \right> = {\mathcal{M}^{(n)}_{t N}}({\eta_{_N}})_{AB}\, \left< {\psi_A^*}\,\psi_B^{}\right>
\end{equation}
Dropping the incoming radiation from both sides, and following the derivation shown in Eqs.~\ref{eq:NLStokeMuellerElments} to \ref{eq:MuellerNLBoxed}, the ensemble representation of the matrix element becomes:
%%%%% Eq: Stokes Poloarimetry  %%%%%
\begin{equation}\label{eq:MuellerNLEnsemble}
{\mathcal{M}^{(n)}_{t N}} = {\textstyle{1 \over c_\eta}} \sum_{e}^{} p_e 
\left(\chi^{(n)*}_{aA}{\left( {{\tau^{}_t }} \right)_{ab}}{\chi^{(n)}_{bB}}{\left( {{\eta_{_N}}} \right)_{BA}}\right)
\end{equation}
By taking the constants ${\left( {{\tau^{}_t }} \right)_{ab}}$ and ${\left( {{\eta_{_N}}} \right)_{BA}}$ out of the summation, and substituting the relation $ \sum_{e}^{} p_e 
\left(\chi^{(n)*}_{aA}{\chi^{(n)}_{bB}}\right) = \left<\chi^{(n)*}_{aA}{\chi^{(n)}_{bB}}\right>_e$ in Eq.~\ref{eq:MuellerNLEnsemble}, the nonlinear Mueller element for the ensemble becomes:
%%%%% Eq: Stokes Poloarimetry  %%%%%
\begin{equation}\label{eq:MuellerNLEnsemble2}
{\mathcal{M}^{(n)}_{t N}} =  
{\textstyle{1 \over c_\eta}} \left<\chi^{(n)*}_{aA}{\chi^{(n)}_{bB}}\right>_e {\left( {{\tau^{}_t }} \right)_{ab}}{\left( {{\eta_{_N}}} \right)_{BA}}
\end{equation}
where $\left< \right>_e$ stands for the average over the $e$ ensemble. The right-hand side of Eq.~\ref{eq:MuellerNLEnsemble2} has a similar form to Eq.~\ref{eq:MuellerNLBoxed}, except that here an ensemble of $\chi^{(n)}$ are considered (the order of variables is a non-issue because both equations are in the elemental form). The correlation matrix $\rm X$ forming from $\left<\chi^{(n)*}_{aA}{\chi^{(n)}_{bB}}\right>_e$ contains all the information about the ensemble, and in the case of a perfectly homogeneous medium reduces to a single source.

Note, since the generated light is no longer originating from a single source, but rather from an ensemble of sources that may not be necessarily coherent, then the outgoing radiation may not be fully polarized. This result is a desired and provides a better representation of experimental data from a heterogeneous medium.
\subsection{Combining Nonlinear and Linear Optical Elements}
For a setup, containing a nonlinear optical medium followed by a train of linear optical components, the Mueller Stokes formalism can be used to relate vector of incoming radiation to the outgoing  vector of the nonlinear radiation:
%%%%% Eq: Stokes Poloarimetry: Linear Nonliear Combined  %%%%%
\begin{equation}\label{eq:StokeMuellerLinNonlin}
s'(\omega_\sigma) = {M_t} \cdots {M_1}\mathcal{M}^{(n)}\,S(\omega_1,\omega_2,\cdots,\omega_n )
\end{equation}
where $M_1...M_t$ are the $4\times 4$ linear Mueller matrices that characterize the linear interactions, and $\mathcal{M}^{(n)}$ is the $4\times 9$ for the second-order matrix, $4\times 16$ for the third-order matrix, and $4\times (n+1)^2$ for the $n^\text{th}$ order nonlinear interaction. Therefore, linear and nonlinear Stokes Mueller formalism can be appropriately combined. As an example, we point to derivation of the so-called polarization-in polarization-out ``PIPO" equation for SHG using the double Stokes Mueller polarimetry~\cite{Samim2015_Doubl}. Similar relation also exist for THG intensity equation, which we will show in an upcoming publication.

\section{Conclusion}
The general formalism for nonlinear Stokes Mueller polarimetry is derived. The derivation stems from the basic nonlinear relationship between the polarization density and the resultant outgoing electric field from an intervening material due to the incoming radiation. In nonlinear polarimetry all three components of the expression including the incoming radiation, the material under study, as well as the outgoing radiation, are characterized by real-valued parameters. The state of the incoming radiations is characterized by $(n+1)^2 \times1$ vector; the sample is represented by a $4\times (n+1)^2$ matrix; and the outgoing radiation is simply determined by a conventional $4\times1$ Stokes vector. States are described in terms of electric fields, and the conventional Stokes vectors. The nonlinear matrix $\mathcal{M}^{(n)}$ is derived in terms of nonlinear susceptibilities. The theoretical framework is comprehensive (since it encapsulates all aspects of the polarization state for the outgoing radiation) for a given material and an incoming radiation. Previous successful nonlinear polarimetric studies such as polarization-in polarization-out (PIPO) equations are shown to be a particular case of Stokes Mueller nonlinear polarimetry, where linear polarizations are employed in non-birefringent and non-absorbing materials. The theory describes the polarimetry of important two-photon effects such as SHG, SFG and DFG, as well as three-photon effects including THG and CARS. For each case the polarization state of incoming radiations as well as the nonlinear optical properties of the intervening material can be described in terms of measurable polarimetric quantities.

The coherency matrix is constructed from a vector composed of electric fields of the incoming radiation. The expansion of the coherency matrix is facilitated by a set of matrices with unique properties and form the basis for development of the polarization state vector as well as the susceptibility matrix. Elsewhere the $\eta$ matrices are shown to be the generalized matrices for group SU(n+1), where an (n+1)-dimensional quantum system is described by $(n+1) \times (n+1)$ density matrix~\cite{Kimura2003_TheB}. Therefore, these matrices may be used for quantum-mechanical derivation of nonlinear polarimetry. For an $n^\text{th}$-order process the overall formalism is the same. For example, the material matrices for sum-frequency generation (SFG), difference-frequency generation (DFG) and SHG assume similar form. Similarly, the matrix for three-photon-polarimetry shares the same form for THG and CARS processes. It is conceivable that a similar approach can be taken to express the state for various other frequency mixing techniques including two-photon absorption, coherent Stokes Raman scattering (CSRS), stimulated Raman scattering (SRS), and parametric amplification. For each of these techniques the polarization states needs to be expressed in terms of the electric fields that nonlinearly interact and result in the nonlinear polarization density. For higher-order techniques such as fourth and fifth harmonics the corresponding higher dimension $\eta$ matrices may be used.

\bibliography{NLSMBIB}

\begin{thebibliography}{32}%
\makeatletter
\providecommand \@ifxundefined [1]{%
 \@ifx{#1\undefined}
}%
\providecommand \@ifnum [1]{%
 \ifnum #1\expandafter \@firstoftwo
 \else \expandafter \@secondoftwo
 \fi
}%
\providecommand \@ifx [1]{%
 \ifx #1\expandafter \@firstoftwo
 \else \expandafter \@secondoftwo
 \fi
}%
\providecommand \natexlab [1]{#1}%
\providecommand \enquote  [1]{``#1''}%
\providecommand \bibnamefont  [1]{#1}%
\providecommand \bibfnamefont [1]{#1}%
\providecommand \citenamefont [1]{#1}%
\providecommand \href@noop [0]{\@secondoftwo}%
\providecommand \href [0]{\begingroup \@sanitize@url \@href}%
\providecommand \@href[1]{\@@startlink{#1}\@@href}%
\providecommand \@@href[1]{\endgroup#1\@@endlink}%
\providecommand \@sanitize@url [0]{\catcode `\\12\catcode `\$12\catcode
  `\&12\catcode `\#12\catcode `\^12\catcode `\_12\catcode `\%12\relax}%
\providecommand \@@startlink[1]{}%
\providecommand \@@endlink[0]{}%
\providecommand \url  [0]{\begingroup\@sanitize@url \@url }%
\providecommand \@url [1]{\endgroup\@href {#1}{\urlprefix }}%
\providecommand \urlprefix  [0]{URL }%
\providecommand \Eprint [0]{\href }%
\providecommand \doibase [0]{http://dx.doi.org/}%
\providecommand \selectlanguage [0]{\@gobble}%
\providecommand \bibinfo  [0]{\@secondoftwo}%
\providecommand \bibfield  [0]{\@secondoftwo}%
\providecommand \translation [1]{[#1]}%
\providecommand \BibitemOpen [0]{}%
\providecommand \bibitemStop [0]{}%
\providecommand \bibitemNoStop [0]{.\EOS\space}%
\providecommand \EOS [0]{\spacefactor3000\relax}%
\providecommand \BibitemShut  [1]{\csname bibitem#1\endcsname}%
\let\auto@bib@innerbib\@empty
%</preamble>
\bibitem [{\citenamefont {Bickel}\ \emph {et~al.}(1976)\citenamefont {Bickel},
  \citenamefont {Davidson}, \citenamefont {Huffman},\ and\ \citenamefont
  {Kilkson}}]{Bickel1976_Appli}%
  \BibitemOpen
  \bibfield  {author} {\bibinfo {author} {\bibfnamefont {W.~S.}\ \bibnamefont
  {Bickel}}, \bibinfo {author} {\bibfnamefont {J.~F.}\ \bibnamefont
  {Davidson}}, \bibinfo {author} {\bibfnamefont {D.~R.}\ \bibnamefont
  {Huffman}}, \ and\ \bibinfo {author} {\bibfnamefont {R.}~\bibnamefont
  {Kilkson}},\ }\href@noop {} {\bibfield  {journal} {\bibinfo  {journal}
  {Proceedings of the National Academy of Sciences of the United States of
  America}\ }\textbf {\bibinfo {volume} {73}},\ \bibinfo {pages} {486}
  (\bibinfo {year} {1976})}\BibitemShut {NoStop}%
\bibitem [{\citenamefont {Bickel}\ and\ \citenamefont
  {Stafford}(1981)}]{Bickel1981_Polar}%
  \BibitemOpen
  \bibfield  {author} {\bibinfo {author} {\bibfnamefont {W.~S.}\ \bibnamefont
  {Bickel}}\ and\ \bibinfo {author} {\bibfnamefont {M.~E.}\ \bibnamefont
  {Stafford}},\ }\href@noop {} {\bibfield  {journal} {\bibinfo  {journal}
  {Journal of Biological Physics}\ }\textbf {\bibinfo {volume} {9}},\ \bibinfo
  {pages} {53} (\bibinfo {year} {1981})}\BibitemShut {NoStop}%
\bibitem [{\citenamefont {Yao}\ and\ \citenamefont
  {Wang}(1999)}]{Yao1999_Two-d}%
  \BibitemOpen
  \bibfield  {author} {\bibinfo {author} {\bibfnamefont {G.}~\bibnamefont
  {Yao}}\ and\ \bibinfo {author} {\bibfnamefont {L.~V.}\ \bibnamefont {Wang}},\
  }\href@noop {} {\bibfield  {journal} {\bibinfo  {journal} {Optics Letters}\
  }\textbf {\bibinfo {volume} {24}},\ \bibinfo {pages} {537} (\bibinfo {year}
  {1999})}\BibitemShut {NoStop}%
\bibitem [{\citenamefont {Jiao}\ \emph {et~al.}(2000)\citenamefont {Jiao},
  \citenamefont {Yao},\ and\ \citenamefont {Wang}}]{Jiao2000_Depth}%
  \BibitemOpen
  \bibfield  {author} {\bibinfo {author} {\bibfnamefont {S.}~\bibnamefont
  {Jiao}}, \bibinfo {author} {\bibfnamefont {G.}~\bibnamefont {Yao}}, \ and\
  \bibinfo {author} {\bibfnamefont {L.~V.}\ \bibnamefont {Wang}},\ }\href@noop
  {} {\bibfield  {journal} {\bibinfo  {journal} {Applied Optics}\ }\textbf
  {\bibinfo {volume} {39}},\ \bibinfo {pages} {6318} (\bibinfo {year}
  {2000})}\BibitemShut {NoStop}%
\bibitem [{\citenamefont {Ghosh}\ \emph {et~al.}(2008)\citenamefont {Ghosh},
  \citenamefont {Vitkin},\ and\ \citenamefont {Wood}}]{Ghosh2008_Muell}%
  \BibitemOpen
  \bibfield  {author} {\bibinfo {author} {\bibfnamefont {N.}~\bibnamefont
  {Ghosh}}, \bibinfo {author} {\bibfnamefont {I.~A.}\ \bibnamefont {Vitkin}}, \
  and\ \bibinfo {author} {\bibfnamefont {M.~F.~G.}\ \bibnamefont {Wood}},\
  }\href@noop {} {\bibfield  {journal} {\bibinfo  {journal} {Journal of
  Biomedical Optics}\ }\textbf {\bibinfo {volume} {13}},\ \bibinfo {pages}
  {044036} (\bibinfo {year} {2008})}\BibitemShut {NoStop}%
\bibitem [{\citenamefont {Ghosh}\ \emph {et~al.}(2009)\citenamefont {Ghosh},
  \citenamefont {Wood}, \citenamefont {Li}, \citenamefont {Weisel},
  \citenamefont {Wilson}, \citenamefont {Li},\ and\ \citenamefont
  {Vitkin}}]{Ghosh2009_Muell}%
  \BibitemOpen
  \bibfield  {author} {\bibinfo {author} {\bibfnamefont {N.}~\bibnamefont
  {Ghosh}}, \bibinfo {author} {\bibfnamefont {M.~F.}\ \bibnamefont {Wood}},
  \bibinfo {author} {\bibfnamefont {S.~H.}\ \bibnamefont {Li}}, \bibinfo
  {author} {\bibfnamefont {R.~D.}\ \bibnamefont {Weisel}}, \bibinfo {author}
  {\bibfnamefont {B.~C.}\ \bibnamefont {Wilson}}, \bibinfo {author}
  {\bibfnamefont {R.~K.}\ \bibnamefont {Li}}, \ and\ \bibinfo {author}
  {\bibfnamefont {I.~A.}\ \bibnamefont {Vitkin}},\ }\href@noop {} {\bibfield
  {journal} {\bibinfo  {journal} {J Biophotonics}\ }\textbf {\bibinfo {volume}
  {2}},\ \bibinfo {pages} {145} (\bibinfo {year} {2009})}\BibitemShut {NoStop}%
\bibitem [{\citenamefont {Wood}\ \emph {et~al.}(2009)\citenamefont {Wood},
  \citenamefont {Ghosh}, \citenamefont {Moriyama}, \citenamefont {Wilson},\
  and\ \citenamefont {Vitkin}}]{Wood2009_Proof}%
  \BibitemOpen
  \bibfield  {author} {\bibinfo {author} {\bibfnamefont {M.~F.}\ \bibnamefont
  {Wood}}, \bibinfo {author} {\bibfnamefont {N.}~\bibnamefont {Ghosh}},
  \bibinfo {author} {\bibfnamefont {E.~H.}\ \bibnamefont {Moriyama}}, \bibinfo
  {author} {\bibfnamefont {B.~C.}\ \bibnamefont {Wilson}}, \ and\ \bibinfo
  {author} {\bibfnamefont {I.~A.}\ \bibnamefont {Vitkin}},\ }\href@noop {}
  {\bibfield  {journal} {\bibinfo  {journal} {J Biomed Opt}\ }\textbf {\bibinfo
  {volume} {14}},\ \bibinfo {pages} {014029} (\bibinfo {year}
  {2009})}\BibitemShut {NoStop}%
\bibitem [{\citenamefont {Campagnola}\ \emph {et~al.}(1999)\citenamefont
  {Campagnola}, \citenamefont {Wei}, \citenamefont {Lewis},\ and\ \citenamefont
  {Loew}}]{Campagnola1999_High-}%
  \BibitemOpen
  \bibfield  {author} {\bibinfo {author} {\bibfnamefont {P.~J.}\ \bibnamefont
  {Campagnola}}, \bibinfo {author} {\bibfnamefont {M.~D.}\ \bibnamefont {Wei}},
  \bibinfo {author} {\bibfnamefont {A.}~\bibnamefont {Lewis}}, \ and\ \bibinfo
  {author} {\bibfnamefont {L.~M.}\ \bibnamefont {Loew}},\ }\href@noop {}
  {\bibfield  {journal} {\bibinfo  {journal} {Biophysical Journal}\ }\textbf
  {\bibinfo {volume} {77}},\ \bibinfo {pages} {3341} (\bibinfo {year}
  {1999})}\BibitemShut {NoStop}%
\bibitem [{\citenamefont {Campagnola}\ \emph {et~al.}(2006)\citenamefont
  {Campagnola}, \citenamefont {Mohler}, \citenamefont {Plotnikov},\ and\
  \citenamefont {Millard}}]{Campagnola2006_Secon}%
  \BibitemOpen
  \bibfield  {author} {\bibinfo {author} {\bibfnamefont {P.~J.}\ \bibnamefont
  {Campagnola}}, \bibinfo {author} {\bibfnamefont {W.~H.}\ \bibnamefont
  {Mohler}}, \bibinfo {author} {\bibfnamefont {S.}~\bibnamefont {Plotnikov}}, \
  and\ \bibinfo {author} {\bibfnamefont {A.~C.}\ \bibnamefont {Millard}},\
  }\href@noop {} {\bibfield  {journal} {\bibinfo  {journal} {Multiphoton
  Microscopy in the Biomedical Sciences VI}\ }\textbf {\bibinfo {volume}
  {6089}},\ \bibinfo {pages} {C891} (\bibinfo {year} {2006})}\BibitemShut
  {NoStop}%
\bibitem [{\citenamefont {Nucciotti}\ \emph {et~al.}(2010)\citenamefont
  {Nucciotti}, \citenamefont {Stringari}, \citenamefont {Sacconi},
  \citenamefont {Vanzi}, \citenamefont {Fusi}, \citenamefont {Linari},
  \citenamefont {Piazzesi}, \citenamefont {Lombardi},\ and\ \citenamefont
  {Pavone}}]{Nucciotti2010_Probi}%
  \BibitemOpen
  \bibfield  {author} {\bibinfo {author} {\bibfnamefont {V.}~\bibnamefont
  {Nucciotti}}, \bibinfo {author} {\bibfnamefont {C.}~\bibnamefont
  {Stringari}}, \bibinfo {author} {\bibfnamefont {L.}~\bibnamefont {Sacconi}},
  \bibinfo {author} {\bibfnamefont {F.}~\bibnamefont {Vanzi}}, \bibinfo
  {author} {\bibfnamefont {L.}~\bibnamefont {Fusi}}, \bibinfo {author}
  {\bibfnamefont {M.}~\bibnamefont {Linari}}, \bibinfo {author} {\bibfnamefont
  {G.}~\bibnamefont {Piazzesi}}, \bibinfo {author} {\bibfnamefont
  {V.}~\bibnamefont {Lombardi}}, \ and\ \bibinfo {author} {\bibfnamefont
  {F.~S.}\ \bibnamefont {Pavone}},\ }\href@noop {} {\bibfield  {journal}
  {\bibinfo  {journal} {Proceedings of the National Academy of Sciences}\
  }\textbf {\bibinfo {volume} {107}},\ \bibinfo {pages} {7763} (\bibinfo {year}
  {2010})}\BibitemShut {NoStop}%
\bibitem [{\citenamefont {Bancelin}\ \emph {et~al.}(2014)\citenamefont
  {Bancelin}, \citenamefont {Aim\'e}, \citenamefont {Gusachenko}, \citenamefont
  {Kowalczuk}, \citenamefont {Latour}, \citenamefont {Coradin},\ and\
  \citenamefont {Schanne-Klein}}]{Bancelin2014_Deter}%
  \BibitemOpen
  \bibfield  {author} {\bibinfo {author} {\bibfnamefont {S.}~\bibnamefont
  {Bancelin}}, \bibinfo {author} {\bibfnamefont {C.}~\bibnamefont {Aim\'e}},
  \bibinfo {author} {\bibfnamefont {I.}~\bibnamefont {Gusachenko}}, \bibinfo
  {author} {\bibfnamefont {L.}~\bibnamefont {Kowalczuk}}, \bibinfo {author}
  {\bibfnamefont {G.}~\bibnamefont {Latour}}, \bibinfo {author} {\bibfnamefont
  {T.}~\bibnamefont {Coradin}}, \ and\ \bibinfo {author} {\bibfnamefont
  {M.-C.}\ \bibnamefont {Schanne-Klein}},\ }\href@noop {} {\bibfield  {journal}
  {\bibinfo  {journal} {Nature Communications}\ }\textbf {\bibinfo {volume}
  {5}} (\bibinfo {year} {2014})}\BibitemShut {NoStop}%
\bibitem [{\citenamefont {Boulesteix}\ \emph {et~al.}(2004)\citenamefont
  {Boulesteix}, \citenamefont {Beaurepaire}, \citenamefont {Sauviat},\ and\
  \citenamefont {Schanne-Klein}}]{Boulesteix2004_Secon}%
  \BibitemOpen
  \bibfield  {author} {\bibinfo {author} {\bibfnamefont {T.}~\bibnamefont
  {Boulesteix}}, \bibinfo {author} {\bibfnamefont {E.}~\bibnamefont
  {Beaurepaire}}, \bibinfo {author} {\bibfnamefont {M.~P.}\ \bibnamefont
  {Sauviat}}, \ and\ \bibinfo {author} {\bibfnamefont {M.~C.}\ \bibnamefont
  {Schanne-Klein}},\ }\href@noop {} {\bibfield  {journal} {\bibinfo  {journal}
  {Optics Letters}\ }\textbf {\bibinfo {volume} {29}},\ \bibinfo {pages} {2031}
  (\bibinfo {year} {2004})}\BibitemShut {NoStop}%
\bibitem [{\citenamefont {Samim}\ \emph {et~al.}(2014)\citenamefont {Samim},
  \citenamefont {Prent}, \citenamefont {Dicenzo}, \citenamefont {Stewart},\
  and\ \citenamefont {Barzda}}]{Samim2014_Secon}%
  \BibitemOpen
  \bibfield  {author} {\bibinfo {author} {\bibfnamefont {M.}~\bibnamefont
  {Samim}}, \bibinfo {author} {\bibfnamefont {N.}~\bibnamefont {Prent}},
  \bibinfo {author} {\bibfnamefont {D.}~\bibnamefont {Dicenzo}}, \bibinfo
  {author} {\bibfnamefont {B.}~\bibnamefont {Stewart}}, \ and\ \bibinfo
  {author} {\bibfnamefont {V.}~\bibnamefont {Barzda}},\ }\href@noop {}
  {\bibfield  {journal} {\bibinfo  {journal} {Journal of Biomedical Optics}\
  }\textbf {\bibinfo {volume} {19}},\ \bibinfo {pages} {056005} (\bibinfo
  {year} {2014})}\BibitemShut {NoStop}%
\bibitem [{\citenamefont {Freund}\ and\ \citenamefont
  {Deutsch}(1986)}]{Freund1986_Secon}%
  \BibitemOpen
  \bibfield  {author} {\bibinfo {author} {\bibfnamefont {I.}~\bibnamefont
  {Freund}}\ and\ \bibinfo {author} {\bibfnamefont {M.}~\bibnamefont
  {Deutsch}},\ }\href@noop {} {\bibfield  {journal} {\bibinfo  {journal}
  {Optics Letters}\ }\textbf {\bibinfo {volume} {11}},\ \bibinfo {pages} {94}
  (\bibinfo {year} {1986})}\BibitemShut {NoStop}%
\bibitem [{\citenamefont {Barzda}\ \emph {et~al.}(2004)\citenamefont {Barzda},
  \citenamefont {Greenhalgh}, \citenamefont {der Au}, \citenamefont {Squier},
  \citenamefont {Elmore},\ and\ \citenamefont {van Beek}}]{Barzda2004_Secon}%
  \BibitemOpen
  \bibfield  {author} {\bibinfo {author} {\bibfnamefont {V.}~\bibnamefont
  {Barzda}}, \bibinfo {author} {\bibfnamefont {C.}~\bibnamefont {Greenhalgh}},
  \bibinfo {author} {\bibfnamefont {J.~A.}\ \bibnamefont {der Au}}, \bibinfo
  {author} {\bibfnamefont {J.~A.}\ \bibnamefont {Squier}}, \bibinfo {author}
  {\bibfnamefont {S.}~\bibnamefont {Elmore}}, \ and\ \bibinfo {author}
  {\bibfnamefont {J.~H.}\ \bibnamefont {van Beek}}\ }(\bibinfo  {publisher}
  {SPIE},\ \bibinfo {year} {2004})\ pp.\ \bibinfo {pages} {96--103}\BibitemShut
  {NoStop}%
\bibitem [{\citenamefont {Shurcliff}(1962)}]{Shurcliff1962_Polar}%
  \BibitemOpen
  \bibfield  {author} {\bibinfo {author} {\bibfnamefont {W.~A.}\ \bibnamefont
  {Shurcliff}},\ }\href@noop {} {\emph {\bibinfo {title} {Polarized Light:
  Production and Use}}}\ (\bibinfo  {publisher} {Harvard University Press},\
  \bibinfo {year} {1962})\BibitemShut {NoStop}%
\bibitem [{\citenamefont {Kliger}\ \emph {et~al.}(1990)\citenamefont {Kliger},
  \citenamefont {Lewis},\ and\ \citenamefont {Randall}}]{Kliger1990_Polar}%
  \BibitemOpen
  \bibfield  {author} {\bibinfo {author} {\bibfnamefont {D.~S.}\ \bibnamefont
  {Kliger}}, \bibinfo {author} {\bibfnamefont {J.~W.}\ \bibnamefont {Lewis}}, \
  and\ \bibinfo {author} {\bibfnamefont {C.~E.}\ \bibnamefont {Randall}},\
  }\href@noop {} {\emph {\bibinfo {title} {Polarized Light in Optics and
  Spectroscopy}}}\ (\bibinfo  {publisher} {Academic Press},\ \bibinfo {year}
  {1990})\BibitemShut {NoStop}%
\bibitem [{\citenamefont {Azzam}\ and\ \citenamefont
  {Bashara}(1977)}]{Azzam1977_Ellip}%
  \BibitemOpen
  \bibfield  {author} {\bibinfo {author} {\bibfnamefont {R.~M.~A.}\
  \bibnamefont {Azzam}}\ and\ \bibinfo {author} {\bibfnamefont {N.~M.}\
  \bibnamefont {Bashara}},\ }\href@noop {} {\emph {\bibinfo {title}
  {Ellipsometry and polarized light}}}\ (\bibinfo  {publisher} {North-Holland
  Pub. Co.},\ \bibinfo {year} {1977})\BibitemShut {NoStop}%
\bibitem [{\citenamefont {Lien}\ \emph {et~al.}(2013)\citenamefont {Lien},
  \citenamefont {Tilbury}, \citenamefont {Chen},\ and\ \citenamefont
  {Campagnola}}]{Lien2013_Preci}%
  \BibitemOpen
  \bibfield  {author} {\bibinfo {author} {\bibfnamefont {C.-H.}\ \bibnamefont
  {Lien}}, \bibinfo {author} {\bibfnamefont {K.}~\bibnamefont {Tilbury}},
  \bibinfo {author} {\bibfnamefont {S.-J.}\ \bibnamefont {Chen}}, \ and\
  \bibinfo {author} {\bibfnamefont {P.~J.}\ \bibnamefont {Campagnola}},\
  }\href@noop {} {\bibfield  {journal} {\bibinfo  {journal} {Biomedical Optics
  Express}\ }\textbf {\bibinfo {volume} {4}},\ \bibinfo {pages} {1991}
  (\bibinfo {year} {2013})}\BibitemShut {NoStop}%
\bibitem [{\citenamefont {Mazumder}\ \emph {et~al.}(2014)\citenamefont
  {Mazumder}, \citenamefont {Hu}, \citenamefont {Qiu}, \citenamefont {Foreman},
  \citenamefont {Romero}, \citenamefont {Török},\ and\ \citenamefont
  {Kao}}]{Mazumder2014_Revea}%
  \BibitemOpen
  \bibfield  {author} {\bibinfo {author} {\bibfnamefont {N.}~\bibnamefont
  {Mazumder}}, \bibinfo {author} {\bibfnamefont {C.-W.}\ \bibnamefont {Hu}},
  \bibinfo {author} {\bibfnamefont {J.}~\bibnamefont {Qiu}}, \bibinfo {author}
  {\bibfnamefont {M.~R.}\ \bibnamefont {Foreman}}, \bibinfo {author}
  {\bibfnamefont {C.~M.}\ \bibnamefont {Romero}}, \bibinfo {author}
  {\bibfnamefont {P.}~\bibnamefont {Török}}, \ and\ \bibinfo {author}
  {\bibfnamefont {F.-J.}\ \bibnamefont {Kao}},\ }\href@noop {} {\bibfield
  {journal} {\bibinfo  {journal} {Methods}\ }\textbf {\bibinfo {volume} {66}}
  (\bibinfo {year} {2014})}\BibitemShut {NoStop}%
\bibitem [{\citenamefont {Shi}\ \emph {et~al.}(1994)\citenamefont {Shi},
  \citenamefont {McClain},\ and\ \citenamefont {Harris}}]{Shi1994_Gener}%
  \BibitemOpen
  \bibfield  {author} {\bibinfo {author} {\bibfnamefont {Y.}~\bibnamefont
  {Shi}}, \bibinfo {author} {\bibfnamefont {W.~M.}\ \bibnamefont {McClain}}, \
  and\ \bibinfo {author} {\bibfnamefont {R.~A.}\ \bibnamefont {Harris}},\
  }\href@noop {} {\bibfield  {journal} {\bibinfo  {journal} {Physical Review
  A}\ }\textbf {\bibinfo {volume} {49}},\ \bibinfo {pages} {1999} (\bibinfo
  {year} {1994})}\BibitemShut {NoStop}%
\bibitem [{\citenamefont {Begue}\ \emph {et~al.}(2009)\citenamefont {Begue},
  \citenamefont {Everly}, \citenamefont {Hall}, \citenamefont {Haupert},\ and\
  \citenamefont {Simpson}}]{Begue2009_Nonli2}%
  \BibitemOpen
  \bibfield  {author} {\bibinfo {author} {\bibfnamefont {N.~J.}\ \bibnamefont
  {Begue}}, \bibinfo {author} {\bibfnamefont {R.~M.}\ \bibnamefont {Everly}},
  \bibinfo {author} {\bibfnamefont {V.~J.}\ \bibnamefont {Hall}}, \bibinfo
  {author} {\bibfnamefont {L.}~\bibnamefont {Haupert}}, \ and\ \bibinfo
  {author} {\bibfnamefont {G.~J.}\ \bibnamefont {Simpson}},\ }\href@noop {}
  {\bibfield  {journal} {\bibinfo  {journal} {The Journal of Physical Chemistry
  C}\ }\textbf {\bibinfo {volume} {113}},\ \bibinfo {pages} {10166} (\bibinfo
  {year} {2009})}\BibitemShut {NoStop}%
\bibitem [{\citenamefont {Cloude}(1986)}]{Cloude1986_Group}%
  \BibitemOpen
  \bibfield  {author} {\bibinfo {author} {\bibfnamefont {S.~R.}\ \bibnamefont
  {Cloude}},\ }\href@noop {} {\bibfield  {journal} {\bibinfo  {journal}
  {Optik}\ }\textbf {\bibinfo {volume} {75}},\ \bibinfo {pages} {26} (\bibinfo
  {year} {1986})}\BibitemShut {NoStop}%
\bibitem [{\citenamefont {Gil}(2007)}]{Gil2007_Polar}%
  \BibitemOpen
  \bibfield  {author} {\bibinfo {author} {\bibfnamefont {J.~J.}\ \bibnamefont
  {Gil}},\ }\href@noop {} {\bibfield  {journal} {\bibinfo  {journal} {European
  Physical Journal-Applied Physics}\ }\textbf {\bibinfo {volume} {40}},\
  \bibinfo {pages} {1} (\bibinfo {year} {2007})}\BibitemShut {NoStop}%
\bibitem [{Note1()}]{Note1}%
  \BibitemOpen
  \bibinfo {note} {Thus, any matrix notation of $\chi _{iA}^{(n)}$ (i.e. with
  two indices) represent the contracted notation of $n^\protect \text {th}$
  order nonlinear susceptibilities.}\BibitemShut {Stop}%
\bibitem [{\citenamefont {Samim}\ \emph {et~al.}(2015)\citenamefont {Samim},
  \citenamefont {Krouglov},\ and\ \citenamefont {Barzda}}]{Samim2015_Doubl}%
  \BibitemOpen
  \bibfield  {author} {\bibinfo {author} {\bibfnamefont {M.}~\bibnamefont
  {Samim}}, \bibinfo {author} {\bibfnamefont {S.}~\bibnamefont {Krouglov}}, \
  and\ \bibinfo {author} {\bibfnamefont {V.}~\bibnamefont {Barzda}},\
  }\href@noop {} {\bibfield  {journal} {\bibinfo  {journal} {Journal of the
  Optical Society of America B}\ }\textbf {\bibinfo {volume} {32}},\ \bibinfo
  {pages} {451} (\bibinfo {year} {2015})}\BibitemShut {NoStop}%
\bibitem [{\citenamefont {Born}\ and\ \citenamefont
  {Wolf}(1999)}]{Born1999_Princ}%
  \BibitemOpen
  \bibfield  {author} {\bibinfo {author} {\bibfnamefont {M.}~\bibnamefont
  {Born}}\ and\ \bibinfo {author} {\bibfnamefont {E.}~\bibnamefont {Wolf}},\
  }\href@noop {} {\emph {\bibinfo {title} {Principles of Optics}}},\ \bibinfo
  {edition} {7th}\ ed.\ (\bibinfo  {publisher} {Cambridge University Press},\
  \bibinfo {address} {Cambridge},\ \bibinfo {year} {1999})\BibitemShut
  {NoStop}%
\bibitem [{Note2()}]{Note2}%
  \BibitemOpen
  \bibinfo {note} {For any vector $\psi $ of length $n+1$, let $\rho =\psi
  \otimes \psi ^\protect \dag $: then $n\protect \tmspace +\thinmuskip
  {.1667em}{\left [\protect \text {Tr}(\eta _{1}\protect \tmspace +\thinmuskip
  {.1667em}\rho )\right ]}^2=\DOTSB \sum@ \slimits@ \limits
  _{N=2}^{(n+1)^2}{{\left [\protect \text {Tr}(\eta _{N}^{}\protect \tmspace
  +\thinmuskip {.1667em}\rho )\right ]}^2}$. The set of $\eta $ matrices are
  defined in Section~\ref {sec:eta}}\BibitemShut {NoStop}%
\bibitem [{\citenamefont {Kim}\ \emph {et~al.}(1987)\citenamefont {Kim},
  \citenamefont {Mandel},\ and\ \citenamefont {Wolf}}]{Kim1987_Relat}%
  \BibitemOpen
  \bibfield  {author} {\bibinfo {author} {\bibfnamefont {K.}~\bibnamefont
  {Kim}}, \bibinfo {author} {\bibfnamefont {L.}~\bibnamefont {Mandel}}, \ and\
  \bibinfo {author} {\bibfnamefont {E.}~\bibnamefont {Wolf}},\ }\href@noop {}
  {\bibfield  {journal} {\bibinfo  {journal} {Journal of the Optical Society of
  America A}\ }\textbf {\bibinfo {volume} {4}},\ \bibinfo {pages} {433}
  (\bibinfo {year} {1987})}\BibitemShut {NoStop}%
\bibitem [{Note3()}]{Note3}%
  \BibitemOpen
  \bibinfo {note} {This is to simplify the indices and to conform to a Stokes
  Mueller notation of vector = matrix $\times $ vector. The matrices $\eta '$
  can also be used directly for polarimetry, in which case there will be an
  additional index for the entity representing the incoming radiation as well
  as for the entity representing the medium.}\BibitemShut {Stop}%
\bibitem [{Note4()}]{Note4}%
  \BibitemOpen
  \bibinfo {note} {MS, SK and VB have a manuscript under preparation titled
  ``\protect \textit {Third-harmonic generation Stokes Mueller
  Polarimetry}}\BibitemShut {NoStop}%
\bibitem [{\citenamefont {Kimura}(2003)}]{Kimura2003_TheB}%
  \BibitemOpen
  \bibfield  {author} {\bibinfo {author} {\bibfnamefont {G.}~\bibnamefont
  {Kimura}},\ }\href@noop {} {\bibfield  {journal} {\bibinfo  {journal}
  {Physics Letters A}\ }\textbf {\bibinfo {volume} {314}},\ \bibinfo {pages}
  {339} (\bibinfo {year} {2003})}\BibitemShut {NoStop}%
\end{thebibliography}%

\end{document}